\documentstyle{article}
\title{\bf The Yang-Lee Edge Singularity\\
on Feynman Diagrams
}
\author{ {\it D.A. Johnston}\\
         Dept. of Mathematics\\
         Heriot-Watt University\\
         Riccarton\\
         Edinburgh, EH14 4AS, Scotland
         }
\begin{document}
  \maketitle
                      {\Large
                      \begin{abstract}
%
We investigate the Yang-Lee edge singularity
on non-planar random graphs,
which we consider as the Feynman Diagrams
of various $d=0$ field theories,
in order to determine the value of the edge exponent $\sigma$.

We consider the hard dimer model on 
$\phi^3$ and $\phi^4$ random graphs
to test the universality of the exponent
with respect to coordination number,  
and the Ising model in an external field
to test its temperature independence.
The results here for generic (``thin'') random graphs
provide an
interesting counterpoint to the discussion by Staudacher
of these models
on planar random graphs \cite{0}.
%
                        \end{abstract} }
%
  \thispagestyle{empty}
%
%
  \newpage
%
                  \pagenumbering{arabic}

\section{Introduction}

The work of Yang and Lee \cite{YL1, YL2},
later expanded by various other authors \cite{others}, on the behaviour
of spin models in {\it complex} external fields
has provided an important paradigm for the understanding
of the nature of phase transitions. In brief, Yang and Lee
observed that the partition function of a system above its critical temperature $T_c$
was non-zero throughout some neighbourhood
of the real axis in the complex external field plane. As $T \rightarrow T_{c}+$
the endpoints of loci of zeroes moved in to pinch the real axis, signalling the transition.
When such endpoints occur at non-physical (i.e. complex) external field values they can
be considered as ordinary critical points with an associated edge critical
exponent. The picture was later extended by Fisher to
temperature driven transitions by considering the
analyticity properties of the free energy in the complex temperature plane
\cite{fish}.

A few equations to flesh this out would perhaps not go amiss.
On a finite graph $G_n$ with $n$ vertices the free energy of an Ising-like spin model
can be written as
\begin{equation}
F(G_n,\beta,z) = - n h - \ln \prod_{k=1}^n ( z - z_k (\beta))
\end{equation}  
where the fugacity $z = \exp (-2 h)$, and $h$ is the (possibly complex) external field.
The $z_k (\beta)$ are the Yang-Lee zeroes, which in the thermodynamic limit
form dense sets on curves in the complex $z$ plane. In the infinite
volume limit $n \rightarrow \infty$
the free energy per spin is
\begin{equation}
F(G_{\infty},\beta,z) = - h - \int_{-\pi}^{\pi} d \theta \rho(\beta , \theta) \ln ( z - e^{ i \theta} )
\end{equation}
where $\rho(\beta , \theta)$ is the density of the zeroes, which 
can be shown to appear on the unit circle
in the complex $z$ plane in the Ising case
(the Yang-Lee circle theorem).
For $T>T_c$ or, if one prefers $\beta<\beta_c$, there is a gap with
$\rho(\beta , \theta) = 0$ for $| \theta | < \theta_0$, and at these
edge singularities we have
\begin{equation}
\rho(\beta , \theta) \sim ( \theta - \theta_0 )^{\sigma}
\end{equation}
which defines the Yang-Lee edge exponent $\sigma$. This also
implies $M \sim ( \theta - \theta_0 )^{\sigma}$.
Various finite size scaling relations relate the Yang-Lee
exponent to the other critical exponents \cite{IPZ}
and can be used in numerical determinations of
critical behaviour \cite{enzo}. 

The Yang-Lee circle theorem of \cite{YL1,YL2} guarantees
that the roots of the partition function of the Ising model
on a fixed graph $G_n$ lie on the unit circle, but it does
not guarantee that this should be the case when the partition
function is defined by a sum over some class of
random graphs for each $n$
\begin{equation}
Z_n = \sum_{G_n} Z (G_n)
\label{part}
\end{equation}
where $Z(G_n)$ is the partition function on a given graph
in the class.
This is the case for models of 2D quantum gravity, where the
sum is over planar $\phi^3, \phi^4 \ldots $ random graphs,
and in this paper where we will will consider
a sum over thin, {\it non}-planar random graphs
\footnote{We shall call such graphs ``thin'' random graphs
throughout this paper as they appear as the
scalar limit of the matrix fatgraphs that 
are relevant for discussions of 2D gravity.}. 
The work of Staudacher \cite{0} showed that,
nonetheless, the zeroes did appear on the unit circle
for planar graphs, which has recently been confirmed by
numerical finite-size scaling investigations using both
series expansions and Monte-Carlo simulations by Ambj\o rn et.al. \cite{jan}.

In this paper we will consider a partition function of the form
equ.(\ref{part}) for thin random graphs.
Spin models on such thin random graphs are of interest because
they give a way of investigating mean-field effects
thanks to their tree-like local structure \cite{3}.
The advantage
of using the thin random graphs in such investigations 
over genuine
tree-like structures such as the Bethe lattice \cite{4} is that boundary effects
are absent. The complications of being forced
to consider only sites deep within the lattice 
which occur on the Bethe lattice are thus absent.
The motivation for the current work 
is the calculation of the exponent $\sigma$
for thin random graphs
and the testing of its universality.
The calculation of the edge exponent in the Ising model
context also allows us to check its temperature
independence.
A mean-field value for $\sigma$ (i.e. $1/2$)
would demonstrate that the mean-field nature of critical behaviour on such random
graphs models extended to complex couplings.

We shall use the approach of \cite{2,00} to solve both the hard dimer model
and the Ising model itself in a complex external field.
The requisite ensemble of thin random graphs is generated by
considering the scalar limit of a matrix model. 
In all cases
the partition function of equ.(\ref{part}) is given
by an integral  of the form
\begin{equation}
Z_n \times N_n = {1 \over 2 \pi i} \oint { d \lambda \over
\lambda^{2n + 1}} \int { \prod_i d \phi_i \over 2 \pi \sqrt{\det K}}
\exp (- S ),
\label{part2}
\end{equation}
where $K$ is the inverse of the quadratic part of the action $S$,
the $\phi_i$ are the fields which give 
the appropriate decoration of the graph, and $\lambda$ is the vertex coupling.
The factor $N_n$ counts the number of undecorated graphs in the
class of interest, and generically grows factorially with $n$. 
A given graph appears as a particular ``Feynman diagram'' in the expansion of
equ.(\ref{part2}) and the integration over $\lambda$ picks out graphs with $2n$
vertices. The coupling $\lambda$ is irrelevant for the discussion of critical
behaviour as it may be scaled out of the action and hence any saddle point equations.
In the large $n$ limit the integral in equ.(\ref{part2}) may be evaluated by saddle point methods.
Phase transitions appear when an exchange of dominant saddle points occurs, either
continuously giving a second order transition, or discontinuously giving a first order transition.
The saddle point integrals which appear
are the $d=0$ equivalent of those in instanton and large orders calculations
in field theory \cite{Zinn,Lip,Cole}. 
As we are taking an $n \rightarrow \infty$ limit at the start of our calculations,
we are unable to explicitly calculate zeroes on finite lattices and verify the Yang-Lee
circle theorem. However, we shall see that the end-points of the loci of zeroes
do lie on the unit circle in the complex fugacity plane. 

One important property of the Yang-Lee edge exponent $\sigma$ 
in all the models examined so far is that it is independent of $\beta$
(for $\beta<\beta_c$). It is therefore possible in general to get a quick and dirty
determination of its value for Ising models by taking 
the so-called hard dimer limit, which 
corresponds to $\beta \rightarrow 0, h \rightarrow {i \pi \over 2}$, and has certain simplifying
features compared with the general case.  We 
will calculate the edge exponent for the hard dimer model on
$\phi^3$ and $\phi^4$ random graphs in the next section. A calculation of the exponent
for the Ising model proper in the section which follows this will investigate
whether the temperature independence still holds.

\section{Hard Dimer Models}

The partition function of the hard dimer model 
on a given graph is defined  \cite{gaunt} by
\begin{equation}
\Theta(G_n) = 1 + \sum_{i=1}^{e(G_n)} \theta_n(i) \zeta^i
\end{equation}
where $\zeta$ is a dimer activity and $\theta_n(i)$ is the number of ways of
placing $i$ dimers on the 
$e(G_n)$ edges of the graph $G_n$ such that at most one dimer is attached
to each vertex (hence ``hard''). Precisely this expression appears
in taking the limit $\beta \rightarrow 0, h \rightarrow {i \pi \over 2}$
in the high temperature series for the Ising model
partition function. The role of $h$ in the definition of the
edge singularity exponent is taken by $\zeta$ and we have
\begin{equation}
{d \Theta \over d \zeta} \sim ( \zeta - \zeta_0)^{\sigma}
\end{equation}
where $\Theta$ is the appropriate $n \rightarrow \infty$ limit
of $\Theta(G_n)$. For both planar and thin random graphs
we are interested in a sum
\begin{equation}
\Theta_n = \sum_{G_n} \Theta (G_n) 
\end{equation}
and in \cite{0} the appropriate two matrix integral to
generate the dimer partition function on $\phi^3$ and $\phi^4$ 
planar graphs  was written down.
For thin $\phi^3$ graphs the required action for insertion
into equ.(\ref{part}) is
\begin{equation}
S = \frac{1}{2} \left( x^2 + y^2 \right) - \frac{1}{3} x^3 - \sqrt{\zeta} y x^2.
\label{phi3}
\end{equation}
where the $x$ propagators represent the unoccupied links and
the $y$ propagators the dimers. 
The two vertices are
shown in Fig.1.
The $\sqrt{\zeta}$ weight appears because each end 
of the dimer contributes a $\sqrt{\zeta} y x^2$
vertex. We have scaled out $\lambda$ for clarity.
Similarly for $\phi^4$ graphs we find
\begin{equation} 
S = \frac{1}{2} \left( x^2 + y^2 \right) - \frac{1}{4}x^4 - \sqrt{\zeta} y x^3.
\label{phi4}
\end{equation}
The saddle point equations $\partial S / \partial x = \partial S / \partial y
=0$ for both equs.(\ref{phi3},\ref{phi4}) can be easily solved.
Concentrating on the $\phi^3$ case for simplicity, we find
\begin{equation} 
x = - { 1 \pm  \sqrt{ 1 + 8 \zeta} \over 4 \zeta}, \; \; \; y = {1 - x \over 2 \sqrt{\zeta}}
\end{equation}
so the resulting saddle point action is
\begin{equation}
S = \frac{1}{192 \zeta^3} \left(12 \zeta - 8 \zeta \sqrt{1 + 8 \zeta} + 24 \zeta^2 + 1 - \sqrt{1 + 8 \zeta} \right) .
\end{equation}
We can see that the saddle point solution presents a singularity
at a negative value, $\zeta_0 = - 1/8$. The free
energy is given by the logarithm of the action
to leading order in $1/n$, so 
we would expect an inverse square root divergence
at $\zeta_0$
when we differentiate $\ln S$ twice if $\sigma$
were equal to its mean-field value of $1/2$.
This is, indeed, the case. Writing $\zeta = - 1/8
+ \epsilon$ and expanding we find
\begin{equation}
{\partial^2 \Theta  \over \partial \zeta^2} \sim 
{\partial^2 \ln S \over \partial \zeta^2} \sim { 96 \sqrt{2} \over \sqrt{\epsilon}}.
\label{div}
\end{equation}
As $y$ appears at most quadratically in equ.(\ref{phi3}) an alternative
approach \footnote{Also taken in \cite{0} for the planar case in a matrix model
calculation.} is to integrate it out to get the action
\begin{equation}
S = \frac{1}{2} \left( x^2 + y^2 \right) - \frac{1}{3} x^3 - \frac{1}{2} \zeta x^4
\end{equation}
which has the saddle point solution
\begin{equation}
x = - { 1 \pm  \sqrt{ 1 + 8 \zeta} \over 4 \zeta}
\end{equation}
and displays an identical divergence $\sim { 96 \sqrt{2} / \sqrt{\epsilon}}$
at $\zeta_0 = - 1/8$
to equ.(\ref{div}). The geometrical picture of such an integration on the $y$ variables
is that all the dimers are collapsed to give new $x^4$ vertices and assigned the
appropriate weight. Whichever way the calculation is carried out, the appearance
of the square root divergence confirms that $\sigma=1/2$,
which is the  mean-field
value of the edge exponent.

The universality of the result with respect to the coordination number of
the vertices can be confirmed by solving the saddle point equations for
equ.(\ref{phi4}), which we do not reproduce here, or by integrating out
$y$ to give
\begin{equation}
S = \frac{1}{2} \left( x^2 + y^2 \right) - \frac{1}{4} x^4 - \frac{1}{2} \zeta x^6
\end{equation}
which has the saddle point solution
\begin{equation}
x =  { \sqrt{- 1 \pm  \sqrt{ 1 + 8 \zeta}} \over \sqrt{6} \zeta}
\end{equation}
leading to the saddle point action
\begin{equation} 
S = \frac{1}{432} {( -1 + \sqrt{1 + 12 \zeta} ) ( 24 \zeta + 1 - \sqrt{1 + 12 \zeta}) \over \zeta^2}.
\end{equation}
When expanded around the singularity at $\zeta_0 = - 1 /12$ it also gives
a square root divergence
\begin{equation}
{\partial^2 \Theta  \over \partial \zeta^2} \sim
{\partial^2 \ln S \over \partial \zeta^2} \sim { 36 \sqrt{3} \over \sqrt{\epsilon}}.
\label{div2}
\end{equation}
The exponent $\sigma$ is thus seen to be independent of the coordination number
of the random graphs (three and four
for $\phi^3$ and $\phi^4$ graphs, respectively) on which the dimers are placed.

The hard dimer calculation represents a determination of the edge exponent
at one point on the ($h, \beta$) plane. We now turn to the solution of the 
Ising model in an external field in order to confirm this value for generic
points on the singular $h(\beta)$ line.

\section{The Ising Model in an External Field}

The action for the Ising model on $\phi^3$ graphs in an external field
may be written as
\begin{equation}
S = \frac{1}{2} \left( x^2 + y^2 \right) - c x y - \frac{1}{3} e^h x^3 - \frac{1}{3} e^{-h} y^3 
\end{equation}
where $c = \exp ( - 2 \beta )$. The transition point in the model
is determined by first solving the saddle point equations $\partial S / \partial x 
= \partial S / \partial y = 0$ and then using these solutions to determine
at which point the Hessian of the second partial derivatives is zero \cite{1}. 
This will pick up any continuous transitions that are present.
The net result of these (lengthy) calculations is the following formula
for $h(c)$, the curve in the $h, c$ plane along which the Hessian
is zero
\begin{equation}
\exp(h(c)) = \pm {\sqrt{ 2 c \left( 1 + 18 c^2 - 27 c^4 \pm ( 1 - 9 c^2 ) \sqrt{ 1 - 10 c^2 + 9 c^4} \; \;
\right)} \over 4 c }
\label{crit}
\end{equation}

It is perhaps worthwhile to consider the solution in zero external field at this point
for orientational purposes. In that case we have at high temperatures 
\begin{equation}
x = y  = 1 - c
\end{equation}
which bifurcates in
a mean-field magnetisation transition
at $c=1/3$ to the low temperature
solutions
\begin{eqnarray}
x &=& { 1 + c + \sqrt{1- 2 c - 3 c^2} \over 2 } \nonumber \\
y &=& { 1 + c - \sqrt{1- 2 c - 3 c^2 } \over 2}.
\label{isingsol}
\end{eqnarray}
valid for $c<1/3$. The distinguished role of the zero-field critical point
$c=1/3$ is clear in equ.(\ref{crit}).
$\exp(h(c))$ develops an imaginary part for $c>1/3$ (in the high temperature phase)
and if we plot its modulus as in Fig.2 we can see clearly that $|\exp(h(c))| = 1$
for $c>1/3$. This shows that the endpoints, at least,  of the line of zeroes lie
on the unit circle. Without explicit finite size calculations, or simulations
of the model in a complex field in the style of \cite{jan} we cannot say
that the zeroes lie on the unit circle, but it is reasonable to conjecture
that they do, just as for planar graphs.

Extracting the critical exponent from the magnetisation turns out to be easiest
proposition for the Ising model in a field. The expression for the magnetisation in this case is
\begin{equation}
M(e(h)) = { e^{h} x^3 - {e^{-h} y^3 } \over  e^{h}  x^3 +  e^{-h} y^3 }
\label{MM}
\end{equation}
and it is directly related to the density of zeroes on the unit circle by
\begin{equation}
\rho(\theta) \sim \lim_{\rightarrow 1-} Re \; M  \; \left( r e^{i \theta} \right).
\label{magtheta}
\end{equation}
We proceed by substituting the saddle point solutions for $x,y$ in field
that led to equ.(\ref{crit}) into equ.(\ref{MM}) before taking the limit
in equ.(\ref{magtheta}) above.
The positions of the critical endpoints on the unit circle can be extracted 
by examining the discontinuities in the 
resulting expression, and are given by
\begin{equation}
\theta_0 (c) = \pm \frac{1}{2} \tan^{-1} \left( {(9 c^2 - 1) \sqrt{ 1 - 10 c^2 + 9 c^4} \over 1 + 18 c^2 - 27 c^4} \right)
\label{theta}
\end{equation} 
which is also consistent with equ.(\ref{crit}) for $\exp(h)$.
The endpoints
can be seen to move in to pinch the real axis, $\theta_0 \rightarrow 0 \pm$
as $c \rightarrow 1/3 +$, confirming the Yang-Lee picture of
the transition.

An expansion of the density of zeroes around $\theta_0$ for generic
$c$ still rapidly degenerates into considerable, and not very illuminating, algebraic
complexity. However, fixing $c$ and examining
various points along the critical line in equ.(\ref{crit})
reduces the level of difficulty to that of the hard dimer calculations
in the previous section. For any given value of $c>1/3$, we do indeed find that
$\rho ( \theta ) \sim  ( \theta - \theta_0(c) )^{1/2}$ by using equ.(\ref{magtheta}).
In Fig.3
the density of zeroes is plotted against $\theta - \theta_0(c)$ for $c=1/2$ and the square
root nature of the singularity is evident, as can be confirmed by fitting to
the curve for this and other values of $c$.

\section{Discussion}

Both the hard dimer calculations and those for the full Ising model in an external
field produce the mean-field value for the edge exponent. Given the
body of previous results in \cite{1} showing mean-field behaviour in various models
on random graphs (and the observation in the first of \cite{1}
that the saddle point equations in the mean-field models were identical in content
to the recursion relations use to solve the models on trees) this is no great surprise,
though it does provide the first confirmation that the mean-field critical behaviour
on random graphs extends to critical phenomena at complex couplings.
The inherent simplicity of the saddle point equations for random graphs
made obtaining expressions for the singular behaviour and the position
of the critical endpoints a simple task in the hard dimer model
and also allowed for a demonstration of universality
by examining both $\phi^3$ and $\phi^4$ graphs.
The solution of the Ising model in field was rather more forbidding, 
and we have not presented many of the resulting elephantine equations here,
but it was still possible to give a 
simple demonstration by semi-numerical means
that $\sigma=1/2$ along the critical curve.

We have also seen that for the Ising model
the singular endpoints of the line of zeroes stay on the
unit circle in the complex fugacity plane, 
which provides support for the Yang-Lee circle theorem
with partition functions of the form in equ.(\ref{part}) that involve sums
over thin random graphs.
This naturally leads to the conjecture that
all the Yang-Lee zeroes lie on the unit circle for the 
Ising model on thin random graphs, just as on planar graphs, which could be confirmed
by a finite size scaling analysis for the thin random graph model in the
manner of that carried out in \cite{jan} for planar graphs. Another 
possible extension of the present work would be to consider the Fisher zeroes
in the model by examining the behaviour in the complex temperature plane.

Indeed, a more comprehensive investigation of complex phases
for various spin and vertex models
on both planar and  non-planar random graphs, in the manner of that carried
out by Matveev and Shrock \cite{Sch} for regular lattices, might prove
illuminating. In particular, it would be interesting to see if the 
complex temperature and complex field singularities with atypical
(and lattice dependent) exponents found in \cite{Sch} had their counterparts
in random graph models. 

\bigskip
\clearpage \newpage
\begin{figure}[htb]
\vskip 20.0truecm
\includegraphics{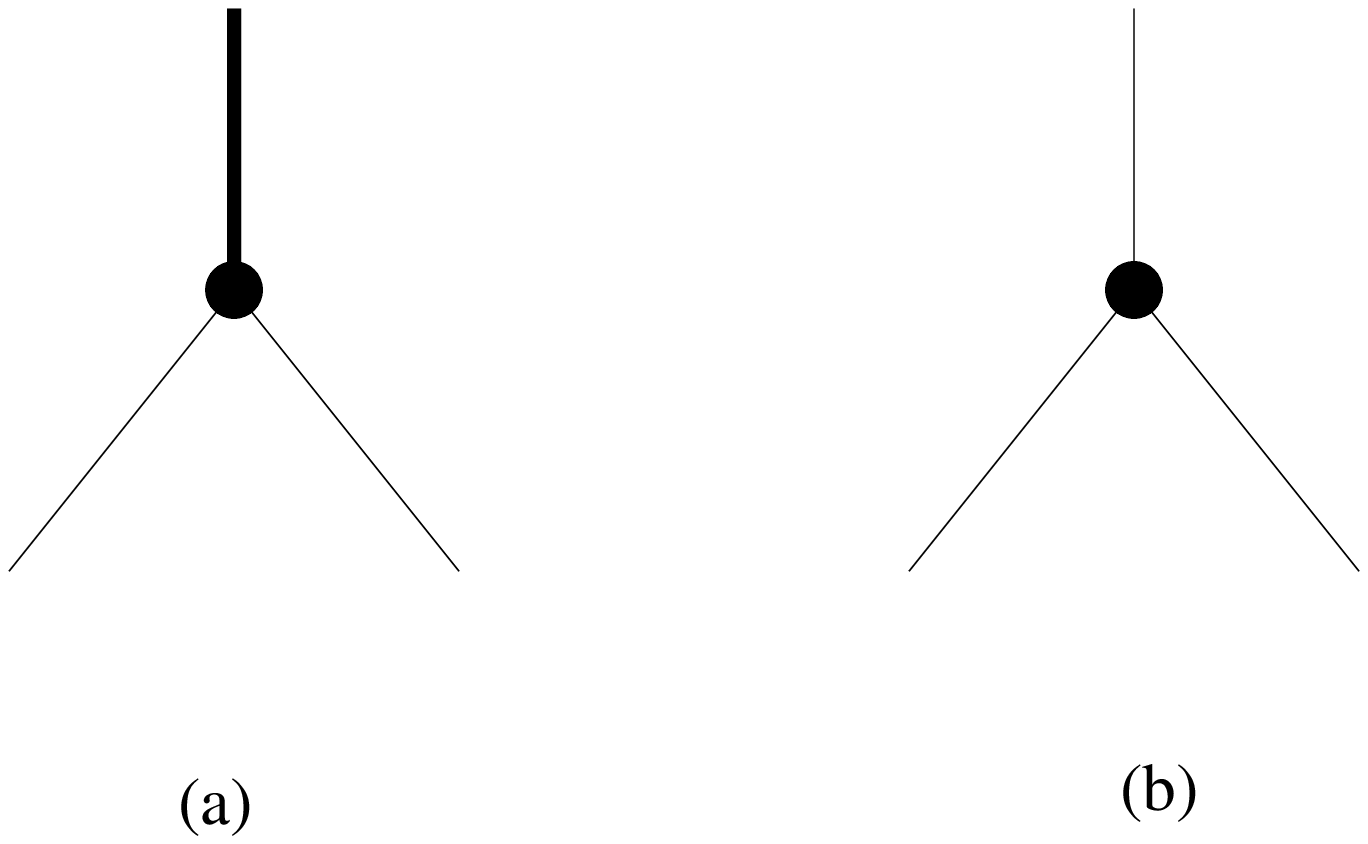}
\caption[]{\label{fig0} The two vertices in the $\phi^3$
dimer model, with the dimer edge drawn in bold. (a) carries a 
weight of $\lambda \sqrt{\zeta}$ and (b) carries a weight
of $\lambda / 3$.}
\end{figure}
\clearpage \newpage
\begin{figure}[htb]
\vskip 20.0truecm
\includegraphics{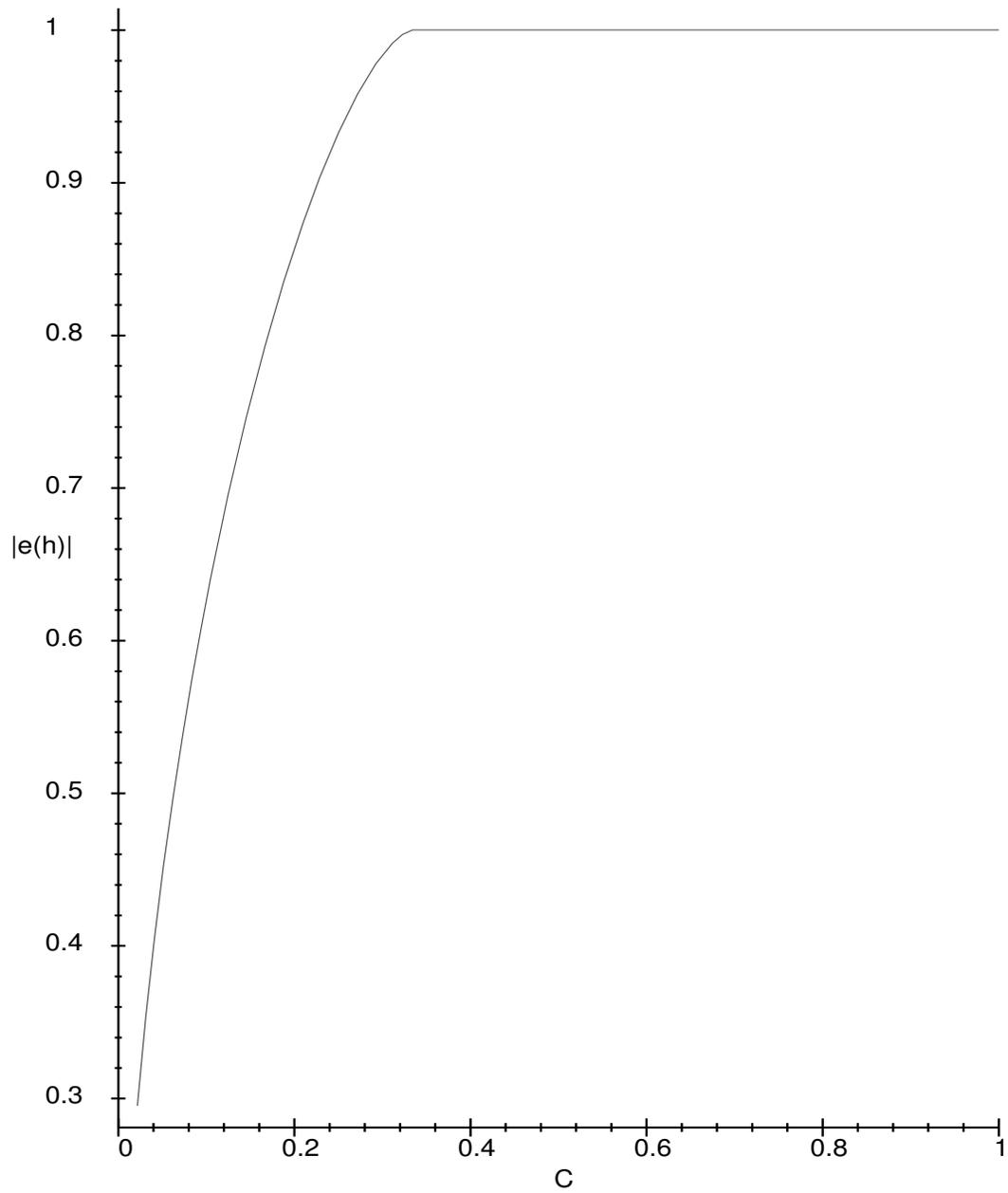}
\caption[]{\label{fig1} The modulus of $\exp(h)$
along the Yang-Lee transition line for $c>1/3$ can be seen to be one}
\end{figure}
\begin{figure}[htb]
\vskip 20.0truecm
\includegraphics{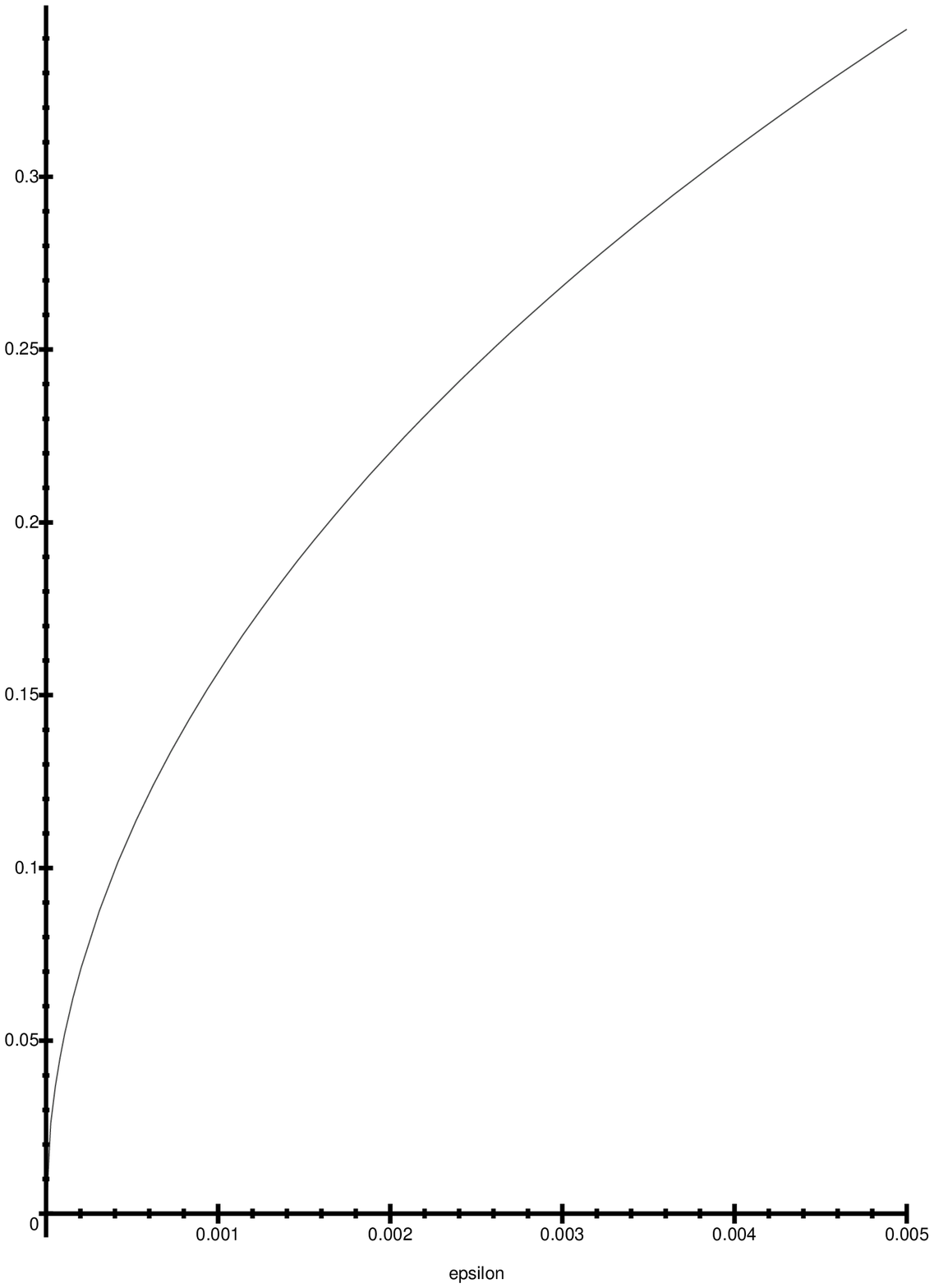}
\caption[]{\label{fig2} The density of zeroes $\rho$ (suitably scaled
for plotting convenience) is plotted against
$\epsilon = \theta - \theta(c)$ for $c=1/2$ to 
show the square root edge singularity.} 
\end{figure}


\begin{thebibliography}{99}
\bibitem{0} M. Staudacher, Nucl. Phys. {\bf  B336} (1990) 349.
\bibitem{YL1} T. D. Lee and C. N. Yang, Phys. Rev {\bf 87} (1952) 410.
\bibitem{YL2} C. N. Yang and T. D. Lee, Phys. Rev. {\bf 87} (1952) 404.
\bibitem{others} J. Lebowitz and O. Penrose, Comm. Math. Phys. {\bf 11} (1968) 99;\\
                 G. Baker, Phys. Rev. Lett. {\bf 20} (1968) 990;\\
                 R. Abe, Prog Theor. Phys. {\bf 37} (1967) 1070; {\it ibid} {\bf 38} (1967) 72;
                 {\it ibid} {\bf 38} (1967) 568;\\
                 S. Ono, Y. Karaki, M. Suzuki and C. Kawabata, J. Phys. Soc. Japan {\bf 25} (1968) 54;\\
                 D. Gaunt and G. Baker, Phys. Rev. {\bf B1} (1970) 1184;\\
                 P. Kortman and R. Griffiths, Phys. Rev. Lett. {\bf 27} (1971) 1439;\\
                 M. Fisher, Phys. Rev. Lett. {\bf 40} (1978) 1611;\\
                 D. Kurtze and M. Fisher, Phys. Rev. {\bf B20} (1979) 2785.
\bibitem{fish} M. Fisher, in ``Lectures in Theoretical Physics'' {\bf VII C} (University of Colorado Press, Boulder, 1965).
\bibitem{IPZ}    C. Itzykson, R. Pearson and J. Zuber, Nucl. Phys. {\bf B220 [FS8]} (1983) 415.
\bibitem{enzo} M. Falcioni, E. Marinari, M. Paciello, G. Parisi and B. Taglienti, Phys. Lett.
{\bf B102} (1981) 220;\\
 M. Falcioni, E. Marinari, M. Paciello, G. Parisi and B. Taglienti, Phys. Lett. 
{\bf B108} (1982) 331;\\
E. Marinari, Nucl. Phys. {\bf B235 [FS11]} (1984) 123.
\bibitem{jan}  J. Ambj\o rn, K. Anagnostopoulos and U. Magnea, Mod. Phys. Lett. {\bf A12} (1997) 1605;\\
               J. Ambj\o rn, K. Anagnostopoulos and U. Magnea, ``Complex Zeroes of the 2-D Ising Model on 
               Dynamical Random Lattices'', to appear in Nucl. Phys. B (proc. suppl.) Lattice97.
\bibitem{3} B. Bollob\'as, ``Random Graphs'', Academic Press, 1985.
\bibitem{4} H. A. Bethe, Proc. Roy. Soc. {\bf A 150} (1935) 552;\\
            C. Domb, Advan. Phys. {\bf 9} (1960) 145;\\
            T. P. Eggarter, Phys. Rev. {\bf B9} (1974) 2989;\\
            E. Muller-Hartmann and J. Zittartz, Phys. Rev. Lett. {\bf
            33} (1974) 893.
\bibitem{1} D. Johnston and P. Plech\' a\v{c}, J. Phys. {\bf A31} (1998) 475;\\ 
            D. Johnston and P. Plech\' a\v{c}, J. Phys. {\bf A30} (1997) 7349;\\
            C. Baillie, D. Johnston and J-P. Kownacki, Nucl. Phys. {\bf B432} (1994) 551;\\
            C. Baillie, W. Janke, D. Johnston and P. Plech\' a\v{c}, Nucl. Phys. {\bf B450}
(1995) 730;\\
            C. Baillie and D. Johnston, Nucl. Phys. {\bf B47} (Proc. Suppl.) (1996) 649;\\
            C. Baillie, N. Dorey, W. Janke and D. Johnston, Phys. Lett {\bf B369} (1996) 123.
\bibitem{2} C. Bachas, C. de Calan and P. Petropoulos, J. Phys. {\bf A27} 
            (1994) 6121.
\bibitem{00} P. Whittle, Adv. Appl. Prob. {\bf 24} (1992) 455;\\
                  J. Stat. Phys. {\bf 56} (1989) 499;\\
                 in {\it Disorder in Physical Systems}, ed. G.R. Grimmett and D.Welsh, (1990) 337.
         
\bibitem{Zinn} E. Brezin, J. Le Guillou and J. Zinn-Justin,
       Phys. Rev. {\bf D15} (1977) 1544;{\it ibid}
1558;\\
G. Parisi, Phys. Lett. {\bf 66B} (1977) 167.
\bibitem{Lip} N. Lipatov, JETP Lett. {\bf 24} (1976) 157;
Sov. Phys. JETP {\bf 44} (1976) 1055; JETP Lett. {\bf 25} (1977) 104;
Sov. Phys. JETP {\bf 45} (1977) 216.
\bibitem{Cole} S. Coleman, Phys. Rev. {\bf D15} (1977) 2929;\\
               C. Callan and S. Coleman, Phys. Rev. {\bf D16} (1977)
1762.
\bibitem{gaunt} D. Gaunt, Phys. Rev. {\bf 179} (1969) 174.
\bibitem{Sch} V. Matveev and R. Shrock, J. Phys. {\bf A28} (1995) 1557;
{\it ibid} 4859;
{\it ibid} 5325; {\it ibid} L533; 
Phys. Lett. {\bf A204} (1995) 353;  J. Phys. {\bf A29} (1996) 803;
Phys. Rev. {\bf E53} (1996) 254; Phys. Lett. {\bf A215}
(1996) 271; Phys. Rev. {\bf E54} (1996) 6174; Phys. Lett. {\bf A221} (1996) 343.  
 
\end{thebibliography}
\end{document}